# Adding salt to expand voltage window of humid ionic liquids

Ming Chen[1,#], Jiedu Wu[2,#], Ting Ye[1], Jinyu Ye[2], Sheng Bi[1], Jiawei Yan[2,*], Bingwei Mao[2], and Guang Feng[1,*]

**Humid hydrophobic ionic liquids, widely used as electrolytes, have narrowed electrochemical windows, because their water, absorbed on the electrode surface, gets involved in electrolysis. In this work, we performed molecular dynamics simulations to explore effects of adding Li-salt in humid ionic liquids on the water adsorbed on the electrode surface. Results reveal that most of water molecules are pushed away from both cathode and anode, by adding salt. The water remained on the electrode is almost bound with Li$^+$, which has significantly lowered activity. The Li$^+$-bonding and re-arrangement of the surface-adsorbed water both facilitate the inhibition of water electrolysis, and thus prevent the reduction of electrochemical windows of humid hydrophobic ionic liquids. This finding is testified by cyclic voltammetry measurements where salt-in-humid ionic liquids exhibit enhanced electrochemical windows. Our work provides the underlying mechanism and a simple but practical approach for protection of humid ionic liquids from performance degradation.**

[1]State Key Laboratory of Coal Combustion, School of Energy and Power Engineering, Huazhong University of Science and Technology (HUST), Wuhan 430074, China. [2]State Key Laboratory of Physical Chemistry of Solid Surfaces, and Department of Chemistry, College of Chemistry and Chemical Engineering, Xiamen University, Xiamen, 361005, China. [#]These authors contributed equally. [*]Correspondence to G.F. (gfeng@hust.edu.cn), J.Y. (jwyan@xmu.edu.cn)

Driven by the demand for the efficient use of intermittent renewable energies and electrical vehicles, the development of electrochemical energy storage (EES) devices with outstanding performance has been at the forefront of energy technologies.[1-4] Nevertheless, developing the safe EES devices with excellent energy density and high power density remains a significant challenge.[1-4] A momentous component for determining the performance of the EES devices is the electrolyte. Room temperature ionic liquid (RTILs), with unique characteristics including excellent thermal stability, nonflammability and especially wide electrochemical windows, are an emerging class of candidates for EES devices,[5-9] such as supercapacitors,[8,9] batteries,[10] and solar cells.[11] However, owing to their hygroscopic nature, RTILs can spontaneously adsorb water from the humid environment, regardless of their hydrophilicity/hydrophobicity.[12] The effect of RTIL-contained water on the electrolyte-electrode interfaces has attracted much attention, since the system performance is controlled by those electrochemical interfaces. For humid RTILs, it has been found that water molecules prefer to accumulate on charged electrode surfaces.[13-15] Very unfavorable to EES devices, the surface-adsorbed water can effectively narrow the electrochemical windows of RTILs due to its electrolysis, resulting in performance degradation.[13,16]

Much of current research has demonstrated that the water electrosorption on the electrode is governed by the working voltage and the association of water molecules with their neighbors (including surface charges, electrode materials and RTIL ions).[14,17,18] It has been reported that owing to the strong interaction with anions, water molecules are excluded from the negatively charged electrode with hydrophilic RTILs.[18] However, it is still an unaddressed issue that, for hydrophobic ionic liquids which have been widely used electrolytes,[19,20] they become humid when exposed to the atmosphere,[12] and unfortunately, their water much favors adsorption on both negatively and positively charged electrodes.[15,16,18] To avoid compromising their wide electrochemical windows, therefore, it is of great importance to minimize surface-adsorbed water and its impact, and thus enhance the voltage window in humid hydrophobic RTILs.

Recently, in battery community, the electrochemical window of aqueous $Li^+$-ion electrolytes is noticeably enlarged by the "water-in-salt" electrolyte.[21-26] This is because the activity of water at the electrode surface decreases significantly due to the strong interaction with $Li^+$ ion that leads to a large shrinkage of "free" water (which is not bound with $Li^+$).[21-27] Some theorical work also revealed that the water molecules can be expulsed from positive electrode with Li$^+$, suppressing the oxygen evolution.[27,28] Herein, analogous to the "water-in-salt" electrolyte, we investigated the effect of adding Li-salt in humid hydrophobic RTILs on the ion and water distributions near electrodes, using molecular dynamics (MD) simulation. The simulation system consists of a slab of humid RTILs confined between two graphite electrode surfaces (**Supplementary Figure 1**). Two hydrophobic RTILs (1-methyl-1-propylpyrrolidinium bis(trifluoromethylsulfonyl)imide, [Pyr$_{13}$][TFSI], and 1-butyl-3-methylimidazolium TFSI, [Bmim][TFSI]) were taken with the salt of Li[TFSI]. Our simulations found that the added salt could expand the electrochemical window of humid hydrophobic RTILs. This finding was testified by our cyclic voltammetry (CV) measurements. We further delve into the molecular structure and intramolecular interaction of salt-in-humid RTILs to understand the mechanism of the electrochemical window expansion, with the help of MD simulations and density functional theory (DFT) calculations as well as infrared spectroscopy (IR).

## Results

### Ion and water distributions

We begin our work by dissecting the electrical double layer (EDL) structure at the electrolyte-electrode interface. **Figure 1** illustrates the number density profiles of ions (Li$^+$, [Pyr$_{13}$]$^+$, and [TFSI]$^-$) and water molecules as a function of distance from the electrode surface with respect to the EDL potential, $\Phi_{EDL}$, which is defined as the potential across the EDL relative to the potential of zero charge (PZC). By comparing EDL structures in humid RTILs before and after adding Li-salt, it can be seen that RTIL cations and anions form alternating layers extending up to few nanometers from the electrode surfaces, and the added salt seems to have weak influence on their number densities (**Supplementary Figures 3-4**). However, the added Li$^+$ has a quite pronounced impact on the water distribution, in particular, that in the interfacial region (0 – 0.35 nm) which is contact-adsorbed on the electrode surface. Panels c and f in **Fig. 1** exhibit water distributuions with varying EDL potential at humid [Pyr$_{13}$][TFSI]/electrode interface and salt-in-humid [Pyr$_{13}$][TFSI]/electrode interface, respectively. These results reveal that in humid [Pyr$_{13}$][TFSI], water preferentially accumulates at the charged electrode surface, consistent with previous studies of the water-in-RTIL mixtures.[14,18,29] In contrast, for salt-in-humid [Pyr$_{13}$][TFSI] electrolytes, most of water molecules are found to be notably excluded from the electrode surface, in particular, showing a complete depletion of water under some negative polarizations and a large removal of water under high polarizations. Further details of the distributions of water molecules and Li$^+$ ions, as well as their development with working voltage, are shown in **Supplementary Figures 5-6**.

Although there are still some water molecules adsorbed on the electrode surface, especially, under large polarizations, such adsorbed water molecules are found to closely stay with the added Li$^+$ ions (contour lines in **Fig. 1f**). We divide the remained adsorbed water into free and bound states, based on the number density profile of Li$^+$ ions around water molecule (i.e., water is considered to be bound to Li$^+$ within a distance of around 0.25 nm, otherwise it is labeled as free water, see **Supplementary Figure 7**).[21,24,27] We, subsequently, examine how the free and bound water in the interfacial region evolves with different polarizations and adding salt (**Fig 2a**). Switching the electrode potential from PZC to -1.5 V, the electrosorbed water is nearly eliminated, and thus unable to participate in the electrolysis. As the voltage becomes more negatively, owing to the strong electrostatic interaction with the charged electrode, the Li$^+$ ions could overcome the energy barrier by RTIL ion layer[30] and then be attracted to the electrode surface (**Fig. 1f**); meanwhile, more water molecules are driven into the interfacial region, but they are almost bound with Li$^+$, so that free water remains depleted near the negative electrode (middle panel of **Fig. 2a**).

As the EDL potential increases from PZC to +1 V, there is always a layer of water molecules adsorbed on electrode, but they are all bound with Li$^+$ ions localized near to the first anion layer (~ 0.53 nm, **Supplementary Figures 4 and 6**). Our DFT calculations show the highest occupied molecular orbital (HOMO) level of the free water is -8.15 eV, which agrees well with previous work,[31] while the HOMO level of the bound water is much lower (-15.48 eV), raising its oxidation potential, which is because in the bound water the electrons from oxygen atom are donated to the Li$^+$.[22,32] When the voltage is larger than +1 V, Li$^+$ ions are expelled from the electrode surface due to the electrostatic repulsion (**Fig. 1f**), and consequently, most of the water molecules, bound with Li$^+$ ions, are taken away from the electrode, resulting in a sharp decrease for free water (middle panel of **Fig. 2a**). The effect of Li$^+$ on the water distribution has been illustrated in **Fig. 2b**.

Therefore, adding salt into humid RTIL could not only help the water to be repulsed from both the negative and positive electrode surfaces but also make the water remained in the interfacial region become bound, which could both potentially protect the humid RTIL from the electrolysis and thus avoid the reduction of electrochemical windows of RTILs due to the existence of water. To validate this, we carried out CV measurements for pure [Pyr$_{13}$][TFSI], humid [Pyr$_{13}$][TFSI] and salt-in-humid [Pyr$_{13}$][TFSI] with HOPG electrodes. As shown in **Fig. 2c**, when [Pyr$_{13}$][TFSI] becomes humid (4474 ppm water), its electrochemical window is obviously narrowed down, shrinking both cathodic and anodic voltage limits. After adding salt (the molar ratio of salt to water is 1:1), the electrochemical window is clearly widened, close to that of pure [Pyr$_{13}$][TFSI]. The simulation and experiment both demonstrate that adding Li$^+$ salt in humid hydrophobic RTIL [Pyr$_{13}$][TFSI] does expand electrochemical window of humid RTIL. Additional modeling and experiment were also carried out to find the impact of the ratio of Li-salt to water, which show the same trend for the electrosorption water distribution and the electrochemical window (**Supplementary Figures 8-9**).

**Origins of bound water and its lowered activity**

In order to understand the mechanism of the voltage window expansion by adding Li-salt, we first explore the fundamental underlying the Li$^+$-bound water. MD simulations of bulk electrolytes were performed to analyze the vibration stretch of O-H of water molecules, which was deployed as the infrared spectroscopy (IR) reporter to monitor the structural change in humid IL electrolytes with adding salt; the spectrum of pure water was treated as the reference.[22,33,34] Comparison of IR spectra of pure water, humid [Pyr$_{13}$][TFSI], and salt-in-humid [Pyr$_{13}$][TFSI] demonstrates a distinct modification for the characteristic of the O-H bond stretching vibration (**Fig. 3a**). To better delineate the change of the stretching vibration of water, three Lorentz peak functions are taken to decompose the IR spectra.[35] It is common and expected that the predicted vibration position has small deviation from the experimental result,[36] and therefore, we, herein, concentrate on the shift of the stretching vibration rather than the absolute value. The O-H stretching vibration in pure water exhibits a broad IR spectrum in 3200~3500 cm$^{-1}$, corresponding to various hydrogen-bonding environments in water clusters.[22,23,33] In humid RTIL, new peaks at higher wavenumber appear at the expense of the water clusters, in accord with previous work.[37,38] With adding salt, the stretching vibrations occur at even higher frequencies when the salt-water ratio varies from 0 to 0.5 then to 4. The increase of the stretching vibrations is also found in our experiments of humid [Pyr$_{13}$][TFSI] with adding salt (**Supplementary Figure 10**). The blue-shift is attributed to the presence of the significantly destroyed H-bond network between water molecules and between water molecules and RTIL ions; thus the water clusters are dramatically diminished, replaced by the Li$^+$(H$_2$O)$_n$ (n = 1, 2, 3, 4) complexes (**Fig. 3b and Supplementary Figure 11a**), similar with those observed in the "water-in-salt" electrolyte.[22,23,33] Consequently, the bond energy of O-H is enhanced and the strength of O-H is reinforced, evidenced by the phenomenon of blue-shift, resulting in a lower reactivity of water.[39,40] Since all conditions are kept unchanged except the amount of Li$^+$ ions, it is reasonable to ascribe the blue-shift to the Li$^+$ addition. Moreover, the blue-shift trend is validated by our DFT calculations (**Supplementary Figure 11b-g**).

The reduced water clusters (**Supplementary Figure 11a**) and the formed bound water (**Fig. 2a**) could be fundamentally understood by the interaction energy between water and the other components as well as water itself. As presented in **Fig. 3c**, the interaction between water and anions is pronounced in humid RTIL, and the interaction between water and water is comparable with that between water and cation. Therefore, water molecules tend to be separated from water clustering by anions and exhibit weaker hydrogen bonding network compared with that in pure water, in agreement with previous work.[37] As the salt was added into humid RTILs, the interaction between water and Li$^+$ rises to be the strongest, and with keeping adding Li$^+$, the interaction between water and Li$^+$ increases gradually and then gets stabilized at about -116 kJ mol$^{-1}$ (**Fig. 3d**). Very differently, the interaction between water and RTIL ions declines to nearly zero, and even the interaction between water molecules becomes a bit repulsive (**Fig. 3c**). The resultant interaction makes water molecules isolated from each other and associated with Li$^+$, leading to a sharp decrease in the proportion of free water (**Fig. 3d**). Briefly, the added Li$^+$ modified the environment and property of water molecules on the basis of the predominated interaction between water and Li$^+$, suggesting that water molecules prefer to tangle with Li$^+$ ions. Consequently, the H-bond networks between water molecules and between water molecules and RTIL ions are disrupted, and the peak positions of O-H stretching vibrations of water are blue-shifted, indicating the reinforced of O-H and then lowered water activity. Meanwhile, the fraction of free water is largely reduced, since water is bound with Li$^+$. These phenomena arising from the addition of Li-salt are responsible for protection of water from electrolysis.

**Understanding effects of adding salt on interfacial water**

We then focus on the origin of salt effect on water distributions at the electrode, by analyzing the free energy as a function of the distance to the electrode surface under different EDL potentials. The potential of mean force (PMF), which represents the variation of free energy, was evaluated by use of umbrella sampling method.[41] For a free water molecule (**Fig. 4a**), a well-defined minimum of PMF occurs at a distance of 0.31 nm from the graphite electrode at PZC, corresponding to an accumulation of water near the electrode; with applying either negative or positive polarization, such minimum becomes more pronounced, resulting in the accumulation of water at charged electrodes.[14,29] However, the situation is different for $Li^+$ in [Pyr$_{13}$][TFSI] (**Fig. 4b**). A deep potential well appears varying from 0.37 to 0.74 nm at PZC, suggesting that $Li^+$ ions preferentially stay away from the electrode surface.[30] As for negative electrode, though a local minimum occurs at 0.21 nm from the electrode with -8.9 kJ mol$^{-1}$ at -2 V, the lowest potential well shifts to 0.81 nm with -28.1 kJ mol$^{-1}$. Changing the potential more negatively, the minimum of free energy moves further from the electrode. The phenomena that the small $Li^+$ ions are unable to access to the negative electrode could attributed to the fact that $Li^+$ ions prefer to remain in the anion layer far from the electrode (**Supplementary Figures 3 and 5**), since the $Li^+$-anion binding (around -370 kJ mol$^{-1}$) is much stronger than that of [Pyr$_{13}$]$^+$-[TFSI]$^-$ pair.[30] Under positive polarization, the broad potential well changes to a distinct valley at around 0.71 nm in the second ion layer, due to a balance between a strong repulsion to positive electrode and sufficient binding from the first anion layer,[30] corresponding quite well with the $Li^+$ number density profiles.

Then, with adding Li-salt, where could be $Li^+$-bound water? We calculated the free energy of a $Li^+$-bound water molecule, as shown in **Fig. 3c**. Different from the water in [Pyr$_{13}$][TFSI], the location of the first minimum PMF at PZC shifts from 0.39 nm to 0.49 nm. When the EDL potential changes to -2 V, we find a positive potential well near electrode (4.8 kJ mol$^{-1}$ at 0.27 nm), which leads to a metastable adsorption.[18] Meanwhile, the bound water induces a pronounced potential well at 0.85 nm. Accordingly, the water molecules prefer to stay far away from the electrode. On the contrary, if the EDL potential is further enlarged (either -3 V or 3 V), different from the single $Li^+$, the water-associated $Li^+$ can pass through the ionic layer and be adsorbed on the electrode surface.

The PMF curves have demonstrated where water, $Li^+$ and $Li^+$-bound water would be predominantly distributed. Specifically, the humid RTIL exhibits water accumulation at polarized electrodes; with adding salt, water molecules could be largely excluded from the electrode under both negative and positive polarizations, compared with the humid RTILs. This different behavior is certainly arising from the strong association with $Li^+$ ions.

**Mechanism of expanded electrochemical window**

With above analyses, the expansion of electrochemical window could be attributed to that the water is repulsed from the electrode surface and the water remained in the interfacial region becomes less active, both originating from the addition of Li-salt.

To delve deeper into salt effect on activity of interfacial water, we evaluate the behavior of surface-adsorbed water in humid RTIL and in salt-in-humid RTILs at the electrodes. The atom density profile and the orientation distribution of water molecules are analyzed in **Fig. 5a-c**. The location and orientation of water molecules in the EDL are found to be sensitive to applied voltage.[28,42] At negative electrode, the interfacial water in humid RTIL exhibits one sharp peak for oxygen atoms and two peaks for hydrogen atoms. Their dipole orientation (at ~111°) and normal orientation (located at 90°) illustrate that most water tends to be perpendicular to the electrode surface, with hydrogen atom pointing to the surface. Such orientations are prone to the hydrogen evolution reaction.[28] After adding salt, the water is re-arranged: the first peak of hydrogen atoms is shifted away from the electrode surface, from 0.22 to 0.25 nm, although there are still two hydrogen peaks in the interfacial region; the dipole orientation of water shifts to 70°. These phenomena suggest that the water molecule adjusts its orientation, together with its hydrogen atoms pushed away from the electrode, making it more difficult for the hydrogen evolution reaction. Meanwhile, for humid RTILs at positive electrode side, the water adopts a configuration parallel to the electrode surface, evidenced by nearly the same peak location (0.3 nm) of the oxygen and hydrogen atom density profiles and the dipole orientation located at 90°; for water in salt-in-humid RTIL, although the position of hydrogen atoms changes little, the peak location of oxygen atoms shifts away from the electrode surface (0.3 to 0.34 nm), and the dipole orientation is changed to 110°. Such structural re-arrangement facilitates the inhibition of water oxidation at positive polarization.[43] Moreover, MD-obtained O-H spectra in the interfacial region are also found to vary with adding salt (**Fig. 5d**), that is, the peak positions of O-H stretching vibrations of interfacial water in salt-in-humid RTIL are shifted towards higher wavenumber, which makes O-H more stable than without salt.

We further testify the generality of the conclusion that the adding salt expands the electrochemical window of humid hydrophobic RTILs, by combined MD-experiment work of another hydrophobic RTIL [Bmim][TFSI]. Although the cations [Pyr$_{13}$]$^+$ and [Bmim]$^+$ are different in terms of their molecular structures, observations both in simulation and experiment (**Supplementary Figures 12-13**) follow the same trend, where water molecules stay away from the electrodes owing to the presence of Li$^+$ and the electrochemical window is enhanced compared with humid [Bmim][TFSI].

## Discussion

We have investigated the effect of adding salt in humid hydrophobic RTILs on the distributions of ion and water at the electrode surface. Combining MD simulations, DFT calculations and CV measurements, we found that the electrochemical stability window of salt-in-humid RTILs is indeed expanded compared with the humid one. MD simulations and DFT calculations provide a molecular understanding of how and why adding salt could dramatically enhance their electrochemical window of humid RTILs. Such enhancement is ascribed to the water repulsed from the electrode surface and the lowered activity of the water remained in the interfacial region, by adding Li-salt.

The water is found to be strongly associated with Li$^+$. Since Li$^+$ ions mostly prefer stay away from the electrode surface, water is intent to be pulled away from the electrode under both negative and positive polarizations, thus extremely reducing the fraction of free water in the interfacial region. For water molecules still adsorbed on the electrode surface, most of them are bound with Li$^+$, resulting in significantly decreased activity, since the strength of O-H in bound water is reinforced due to the association with Li$^+$. Meanwhile, the added Li$^+$ modifies the arrangement of adsorbed water. Explicitly, the position of hydrogen atom is adjusted, making it difficult for the hydrogen evolution to occur at negative electrode; the oxygen atom is pulled away from the positively charged electrode, protecting water from electrolysis. The lower HOMO level is a complementary mechanism for the thermodynamic manner to improve electrolyte oxidation stability under positive polarization.[22,32]

The findings reported with the concept of salt-in-humid RTILs could extend the comprehension of the preferential species electrosorption, and provide a guideline for minimizing water adsorption and altering the structure and properties of water. The mitigation of the adsorbed water and its impact could improve the practical performance of electrochemical energy storage systems as electrochemical window of humid hydrophobic RTILs can be fully exploited, and may also benefit other applications such as RTIL gating and lubrication.

## Methods

**Molecular dynamics simulation**

Molecular dynamics (MD) simulations were utilized to investigate the effect of added salt on the ion and water distributions in the EDL of humid RTIL. Specifically, we employed MD simulation of hydrophobic RTIL-water-salt mixtures confined between two atomically flat graphite electrode surfaces as shown in **Supplementary Figure 1**. All-atom model was taken for two hydrophobic RTILs ([Pyr$_{13}$][TFSI] and [Bmim][TFSI]);[44,45] the SPC/E model was adapted for water molecules.[46] The OPLS force field was used for Li$^+$ ion;[47] carbon atoms in the electrode were modeled using the force fields in ref 48. The sizes of all simulation systems were chosen as long enough to reproduce the bulk-like state of electrolytes in the central region between two electrodes. The number of the species and the distance between two electrodes are given in **Supplementary Table 1** for all systems.

All simulations were performed in NVT ensemble with the MD package GROMACS.[49] Temperature was controlled through the Nosé-Hoover thermostat[50,51] at 333K with coupling constant of 1.0 ps. A cutoff distance of 1.2 nm was employed for van-der-Waals term *via* direct summation. The long-range electrostatic interactions were computed *via* PME method.[52] An FFT grid spacing of 0.1 nm and cubic interpolation were used to compute the electrostatic interaction in the reciprocal space. A cutoff length of 1.2 nm was adopted in the calculation of electrostatic interactions in the real space. The leapfrog integration algorithm was taken to solve the equations of motion, with a time step of 2 fs. Specifically, in order to ensure an adequate description of the electrode polarization effects in the presence of electrolytes, constant potential method (CPM)[2,53,54] was implemented to allow the fluctuations of the charges on electrode atoms. To guarantee the accuracy, the electrode charges were updated on the fly every simulation step. For each simulation, the MD system was first heated at 700 K for 3 ns and then annealed to 333 K over a period of 2 ns, following by another 10 ns to reach equilibrium. After that, a 20 ns production was performed for analysis. Each case was repeated three times with different initial configurations to certify the accuracy of the simulation results.

**Quantum molecular orbital analysis**

DFT calculations on the molecular orbital of water and water/$Li^+$ complexes were performed with the Gaussian 09 program (revision, D.01).[55] The basis sets implemented in the Gaussian program were used. The geometries of water and water/$Li^+$ complexes were fully optimized using B3LYP/DFT method with 6-311G** basis set. Based on their optimized geometries, the HOMO energy levels were calculated at the same level.

**Experimental materials and measurements**

RTILs [$Pyr_{13}$][TFSI] and [Bmim][TFSI] were purchased from IoLiTec in the highest available quality (99%); LiTFSI (99.95%) was purchased from Sigma-Aldrich. Prior to each measurement, RTILs were purified through ultra-pure water (Milli-Q, 18.2 MΩ·cm) and then vacuum-dried for 24 hours at 80 °C in a glovebox filled with ultrapure Argon (Linde Industrial Gases, 99.999%) to remove water as much as possible, and then used as dry RTILs. Humid RTILs are prepared *via* adding ultra-pure water into RTILs. RTILs-LiTFSI-water mixtures were prepared by adding ultra-pure water and LiTFSI into RTILs, and then the mixtures were stirred up for 12 hours till homogeneous solutions were formed. Water contents were determined by Karl Fischer Coulometer (Metrohm, KF-831).[56] Cyclic voltammetry measurements were carried out in a glovebox by using an Autolab electrochemical workstation (Eco Chemie, The Netherlands). Silver wire and platinum wire were used as reference electrode and counter electrode, respectively. Highly oriented pyrolytic graphite (HOPG) was used as working electrode. A piece of tape was pressed onto the flat surface and then pulled off to form a newly cleaved clean surface for electrochemical measurements. The electrochemical cell used for measurements was a sealed one, which was isolated from the environment, and the CV measurements were typically completed within half an hour, so that the change of water contents was negligible. Fourier-transform infrared spectroscopy (FTIR) measurements were conducted on a Nexus 870 spectrometer (Nicolet) equipped with a liquid-nitrogen-cooled MCT-A detector. An unpolarized IR radiation sequentially passed through two $CaF_2$ windows with a thin-layer solution (25 μm).

The IR transmission spectrum of the RTILs-LiTFSI-water mixture was taken from 1111 to 4000 cm$^{-1}$ with a resolution of 4 cm$^{-1}$ and averaged 100 times.

## Acknowledgements

G.F., M.C., T.Y. and S.B. acknowledge the funding support from the National Natural Science Foundation of China (51876072, 51836003). J.W.Y., B.W.M., J.D.W., and J.Y.Y. acknowledge the support of Natural Science Foundation of China (21673193, 21533006 and 21727807). G.F. also thanks the Fundamental Research Funds for the Central Universities (2019kfyXMBZ040). The computation is completed using Tianhe II supercomputer in National Supercomputing Center in Guangzhou.

## Author contributions

G.F. conceived this research. G.F. and J.W.Y. designed the work of simulation and experiment, respectively. M.C., T.Y. and S.B. carried out all simulations. J.D.W., J.Y.Y., J.W.Y. and B.W.M. carried out the experiment. M.C. drafted the manuscript. All authors contributed to the analysis and discussion of the data and revision of the manuscript.

**Competing financial interests:** The authors declare no competing interests.

# Figure 1

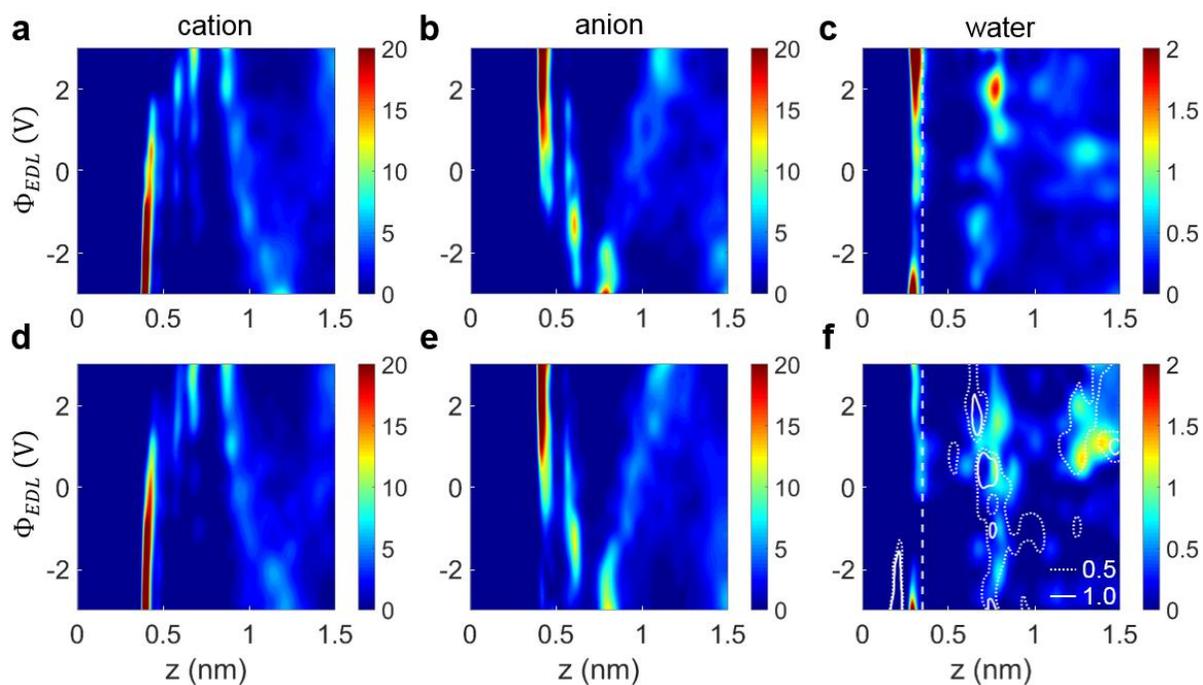

**Fig. 1 | Ion and water distributions under various voltages. a-c,** Number densities of cation (**a**), anion (**b**), and water (**c**) in humid [Pyr$_{13}$][TFSI] as a function of distance from the electrode. **d-f,** The number densities of cation (**d**), anion (**e**), and water as well as Li$^+$ (**f**), in salt-in-humid [Pyr$_{13}$][TFSI]. The vertical dash lines (z = 0.35 nm) in (**c**) and (**f**) represents the right boundary of the interfacial region. The contour in (**f**) indicates number densities of Li$^+$ ions. Unit: # nm$^{-3}$. The molar ratio of Li$^+$ to water is 1:1.

**Figure 2**

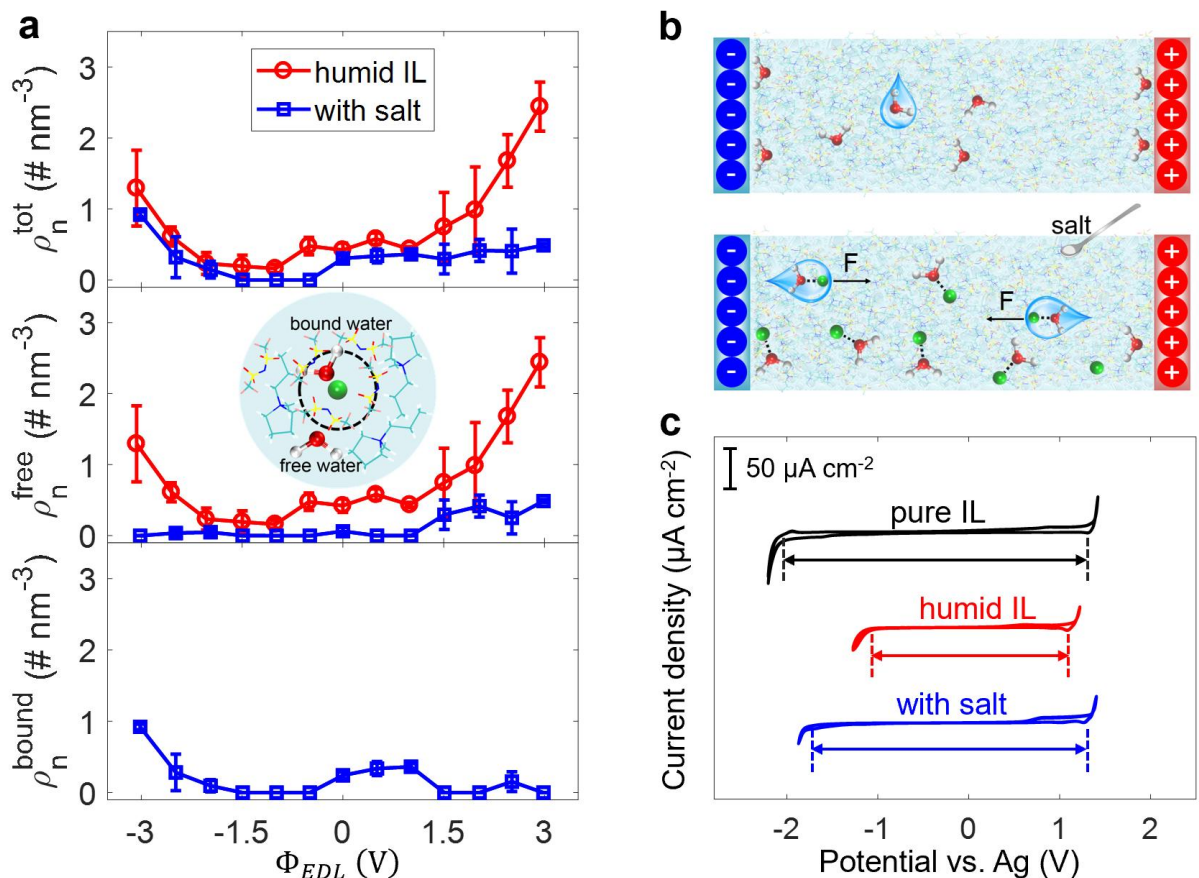

**Fig. 2 | Effect of adding salt on the interfacial water and voltage window. a,** Electrosorption of water from humid [Pyr$_{13}$][TFSI] with/without adding salt. The top, middle and bottom panels are, respectively, the total, free and bound water in the interfacial region. **b,** Schematic of effect of adding salt on water electrosorption. **c,** Cyclic voltammograms of HOPG in pure [Pyr$_{13}$][TFSI] (black line), humid [Pyr$_{13}$][TFSI] (water content: ~4474 ppm, red line) and salt-in-humid [Pyr$_{13}$][TFSI] electrolyte ($n_{Li^+}:n_{water} = 1:1$, blue line). Scan rate: 100 mV s$^{-1}$.

**Figure 3**

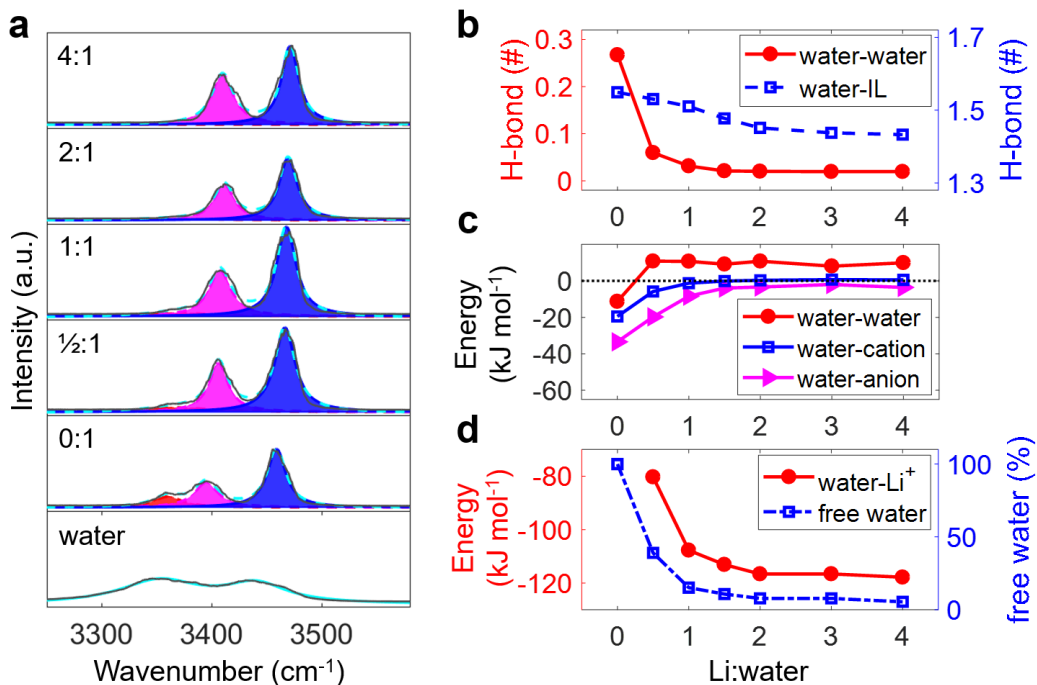

**Fig. 3 | Microscopic understanding of salt effect on water in humid RTILs. a**, MD-calculated IR spectra of water in humid [Pyr$_{13}$][TFSI], as well as humid [Pyr$_{13}$][TFSI] with different salt-water ratio. The gray lines represent the IR spectra. The red, pink, and blue lines are the first, second, and third fitting spectra, respectively. The cyan dash lines represent the summation of the fitting spectra. **b**, H-bonds between water molecules (left axis) and between water molecules and RTIL ions (right axis). **c**, Interaction energy between water and water (red), cation (blue) and anion (pink). **d**, Interaction energy between water and Li$^+$ (left axis), and the proportion of free water as a function of the salt ratio (right axis).

**Figure 4**

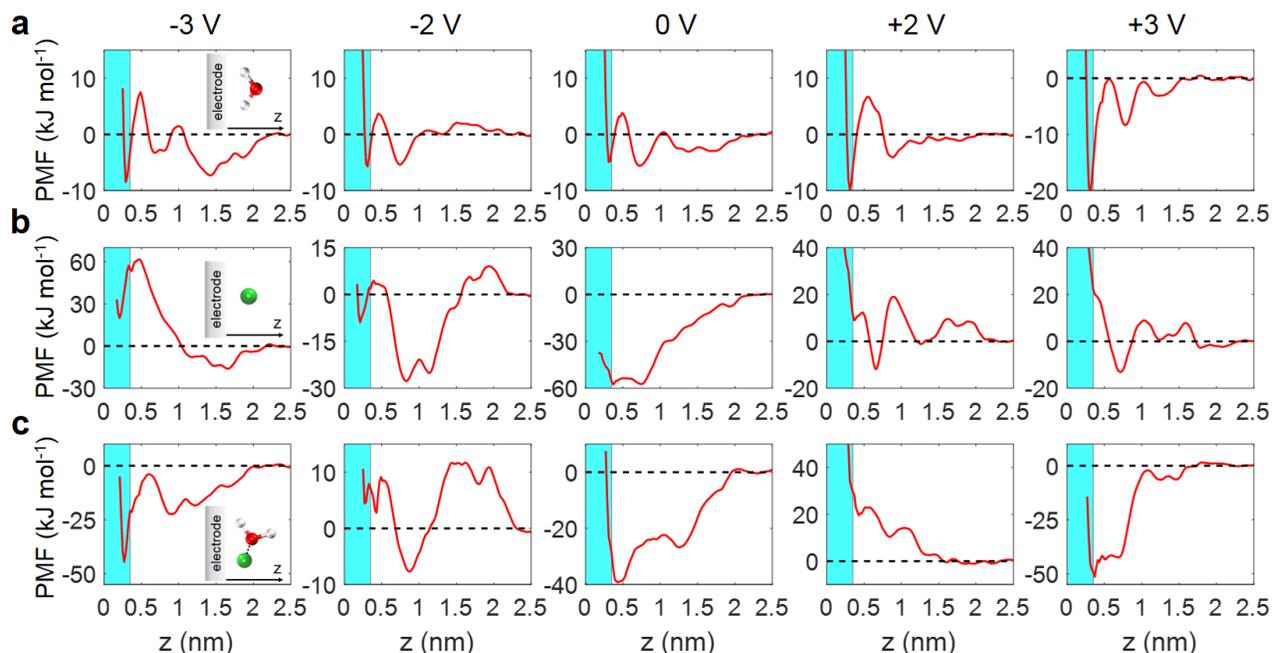

**Fig. 4 | The tendency for electrosorption of water, Li$^+$ and Li$^+$-water pair at electrodes. a-c,** The potential of mean force (PMF) of water **(a)**, Li$^+$ **(b)** and Li$^+$-bound water **(c)** in [Pyr$_{13}$][TFSI] as a function of distance from the electrode. The PMF curves are calculated *via* umbrella sampling along the distance from the electrode surface. The green shaded region (z < 0.35 nm) is considered as the interfacial region.

**Figure 5**

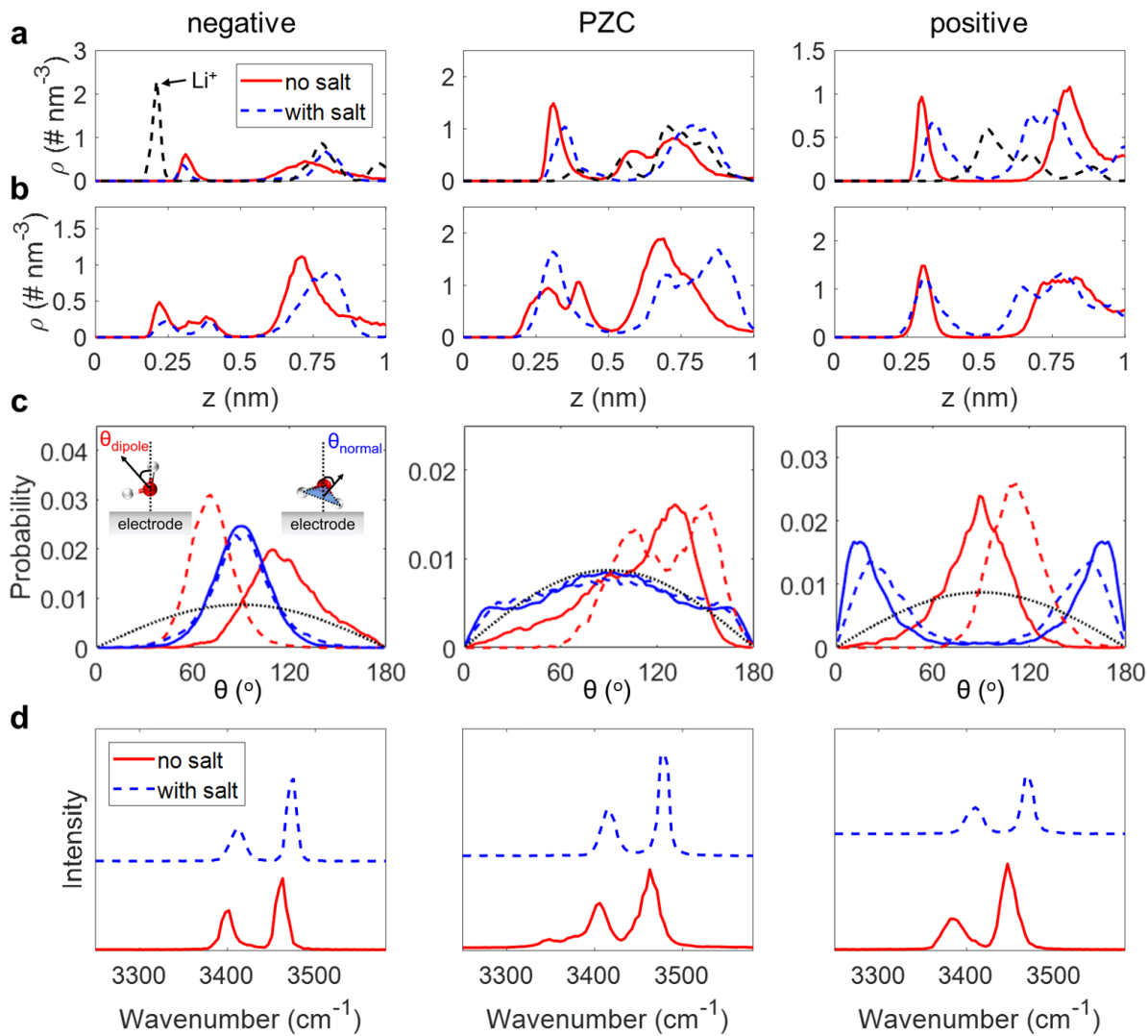

**Fig. 5 | Effect of adding salt on the interfacial water of humid RTIL at the electrified surface. a,** The oxygen atom number densities of water in humid RTIL (red solid line) and in salt-in-humid RTIL (blue dash line) at electrodes. The black dash line is the atom number density of $Li^+$ in salt-in-humid RTIL. **b,** The hydrogen atom number densities of water in humid RTIL (red solid line) and in salt-in-humid RTIL (blue dash line) at electrodes. **c,** The orientation of water in humid RTIL (solid lines) and in salt-in-humid RTIL (dash lines) at electrodes. $\theta_{dipole}$ is defined as the angle between the normal of electrode surface and the water vector, and $\theta_{normal}$ is the angle formed between the normal of electrode surface and the normal of water plane. The red and blue solid lines represent the dipole orientation and normal orientation of water in humid RTIL, respectively. The red and blue dash lines represent the dipole orientation and normal orientation of water in salt-in-humid RTIL, respectively. The black dotted line represents the orientation of bulk water. **d**, IR spectra of interfacial water in humid RTIL (red line) and in salt-in-humid RTIL (blue line). The EDL potentials are -2 V and 1 V for negative electrode and positive electrode, respectively.

20